\documentclass[nofootinbib,aps,11pt]{revtex4}
\usepackage{graphicx}
\usepackage{cancel}
\usepackage{amssymb}
\usepackage{textcomp}
\usepackage{amsmath}
\usepackage{bm}
\usepackage{times}
\usepackage{epsfig}
\usepackage{color}
\usepackage{citesort}

\begin{document}
\preprint{}
\vfill \preprint{} \vspace{2.0cm}
\title{\Large On the Role of Low-Energy CP Violation in Leptogenesis}
\vspace{4.0cm}
\author{Steve Blanchet$^{1}$, Pavel Fileviez P\'erez$^{2}$}
\affiliation{
$^{1}$ Maryland Center for Fundamental Physics and Department of Physics \\
University of Maryland, College Park, Maryland 20742, USA
\\
$^{2}$ University of Wisconsin-Madison, Department of Physics\\
1150 University Avenue, Madison, WI 53706, USA}
\date{\today}
\begin{abstract}
The link between low-energy $C\!P$ violation and leptogenesis became
more accessible with the understanding of flavor effects. However, a
definite well-motivated model where such a link occurs was still
lacking. Adjoint $SU(5)$ is a simple grand unified theory where
neutrino masses are generated through the Type I and Type III seesaw
mechanisms, and the lepton asymmetry is generated by the fermionic
triplet responsible for the Type III seesaw. We focus exclusively on
the case of inverted hierarchy for neutrinos, and we show that
successful flavored leptogenesis in this theory strongly points
towards low-energy $C\!P$ violation. Moreover, since the range of
allowed masses for the triplet is very restricted, we find that the
discovery at the LHC of new states present in the theory, together
with proton decay and unification of gauge couplings, can conspire
to provide a hint in favor of leptogenesis.
\end{abstract}
 \maketitle
\section{Introduction}
One of the main arguments in favor of the search for low-energy
$C\!P$ violation in the lepton sector is that this may provide an
indication about the origin of the matter-antimatter asymmetry in
the Universe. This is due to the connection between
leptogenesis~\cite{Fukugita:1986hr} and neutrino masses within the
seesaw
mechanism~\cite{Minkowski:1977sc,Yanagida,Gell-Mann,Glashow,Barbieri:1979ag,Mohapatra:1980yp}.
However, it turns out that such a link is difficult to establish
without assuming particular structures of the mass matrices in the
model~\cite{Branco:2001pq,Frampton:2002qc}. This was especially true
in the case of ``unflavored leptogenesis'' (see e.g. the recent
review~\cite{Davidson:2008bu} and references therein). But even with
flavored leptogenesis~\cite{Abada:2006fw,Nardi:2006fx}, the link can
be obscure in the most general case~\cite{Davidson:2007va}, if one
does not assume specific values for the parameters in the
high-energy sector. For example, the assumption of subdominant
$C\!P$-violation in the high-energy sector, in which case low-energy
phases in the PMNS matrix are directly related to the size of the
baryon asymmetry through leptogenesis, has recently attracted some
attention~\cite{Blanchet:2006be,Pascoli:2006ie,Pascoli:2006ci,Branco:2006ce,Anisimov:2007mw,Molinaro:2007uv,
Molinaro:2008cw}.

In order to have a connection between leptogenesis and low-energy
$C\!P$ violation, one has to restrict the number of high-energy
phases. This is achieved in the so-called two right-handed (2RH)
neutrino model~\cite{Frampton:2002qc,Ibarra:2003up}, or
$N_3$-decoupling limit~\cite{Chankowski:2003rr}, where only one
high-energy phase is present, instead of three for the model with
three RH neutrinos. With two low-energy phases (the Dirac phase and
one Majorana phase) in the PMNS matrix, it is reasonable to guess
that the low-energy phases can have an important impact. It was
shown in~\cite{Blanchet:2008pw} that, in the case of inverted
hierarchy for neutrinos, the region of successful flavored
leptogenesis was much larger when the low-energy phases were
non-zero. Similar results were presented
in~\cite{Molinaro:2008cw,Molinaro:2008rg}. However, the model with
two RH neutrinos, though simpler from the point of view of the
number of parameters, is more difficult to motivate since in grand
unified theories, in particular $SO(10)$, one has the same number of
RH neutrinos as the number of families.

It is well-known that RH neutrinos and leptogenesis are naturally
embedded in $SO(10)$-based theories.
See~\cite{DiBari:2008mp,Abada:2008gs} for recent studies in this
context. Recently, a realistic grand unified theory based on the
$SU(5)$ gauge symmetry has been proposed~\cite{Perez:2007rm}, where
neutrino masses are generated through a Type~I plus
Type~III~\cite{Foot:1988aq} seesaw mechanism. This case is similar
to the 2RH neutrino model from the point of view of the number of
parameters, but, in order to have viable
unification~\cite{Perez:2008ry}, the fermionic triplet responsible
for the Type III seesaw has to be much lighter than the singlet
responsible for the Type I seesaw, implying different Boltzmann
equations~\cite{Hambye:2003rt}. Therefore, in contrast to the usual
renormalizable $SO(10)$ models where one has a Type~I plus Type~II
seesaw mechanism, and one does not know whether the Higgs triplet or
the fermionic singlet is responsible for leptogenesis, here we know
which field in the theory generates the $B-L$ asymmetry.

Leptogenesis within the model~\cite{Perez:2007rm} was investigated
in detail in~\cite{Blanchet:2008cj}, where constraints on the
parameter space of the theory were derived. In this letter, we study
the particular role played by the low-energy $C\!P$-violating phases
on the generation of asymmetry, and we find that in the case of
inverted hierarchy for neutrinos, $C\!P$ violation at low energy is
a crucial ingredient to have successful leptogenesis. Therefore, in
this well-motivated model and assuming that the spectrum for
neutrinos is inverted, low-energy phases naturally play a dominant
role compared to the unknown high-energy phase. Finally, we show
that the discovery at the LHC of new states present in the theory
can lead to a prediction on the proton decay lifetime from gauge
coupling unification and leptogenesis. It is actually remarkable
that such relations can be obtained in this model.

This letter is organized as follows: In section II we present the
model and its main predictions relevant for leptogenesis. In section
III we discuss the crucial role of the low-energy Dirac and Majorana
phases in leptogenesis. In section IV we discuss the correlation
between the unification constraints, proton decay and leptogenesis.
In the last section we summarize our findings.
\section{Adjoint $SU(5)$ Unification, Neutrino Masses and Leptogenesis}
In the context of renormalizable Adjoint
$SU(5)$~\cite{Perez:2007rm}, neutrino masses are generated through
the Type~I plus Type~III seesaw mechanism. In this context the
matter lives in the $\bf{\bar{5}}$, $\bf{10}$ and $\bf{24}$
representations, while the Higgs sector is composed of $\bf{5_H}$,
$\bf{24_H}$ and $\bf{45_H}$. See \cite{Perez:2008ry,Perez:2008ib}
for the phenomenological and cosmological aspects of this proposal.
The fields responsible for the Type~I and Type~III seesaw are
$\rho_0 \sim (1,1,0) \subset {\bf 24}$ and $\rho_3 \sim (1,3,0)
\subset {\bf 24}$, respectively. Integrating out these fields, the
mass matrix for neutrinos reads
\begin{eqnarray}
M^\nu_{\alpha \beta} & = & \left( \frac{h_{\alpha 1} \ h_{\beta 1}}{M_{\rho_3}} \ +
\ \frac{h_{\alpha 2} \ h_{\beta 2}}
{M_{\rho_0}} \right)v_0^2,
\end{eqnarray}
where $M_{\rho_0}$ and $M_{\rho_3}$ are the masses of the fields
responsible for Type~I and Type~III seesaw, respectively, $h$ is the
Yukawa coupling matrix and $v_0$ the vacuum expectation value of the
Standard Model Higgs. See~\cite{Perez:2008ry} for more details. The
theory predicts one massless neutrino at tree level. Therefore, one
can have either a normal neutrino mass hierarchy: $m_1=0$,
$m_2=\sqrt{\Delta m_{\rm sol}^2}$ and $m_3=\sqrt{\Delta m_{\rm
sol}^2 + \Delta m_{\rm atm}^2}$, or an inverted one: $m_3=0$,
$m_2=\sqrt{\Delta m_{\rm atm}^2}$ and $m_1=\sqrt{\Delta m_{\rm
atm}^2 - \Delta m_{\rm sol}^2}$, where $\Delta m_{\rm sol}^2 \simeq
8 \times 10^{-5}$ eV$^{2}$ and $\Delta m_{\rm atm}^2 \simeq 2.5
\times 10^{-3}$ eV$^{2}$ are the mass-squared differences of solar
and atmospheric neutrino oscillations,
respectively~\cite{GonzalezGarcia:2007ib}.

A convenient parametrization of the $3\times 2$ Yukawa coupling
matrix $h$ is given by~\cite{Casas:2001sr}
\begin{equation}\label{ci}
h=U D_{\nu}^{1/2}\Omega D_{\rho}^{1/2}/v_0,
\end{equation}
where $U$ is the PMNS lepton mixing matrix, $D_{\nu}={\rm
diag}(m_{1},m_2, m_3)$ and $D_{\rho}={\rm
diag}(M_{\rho_3},M_{\rho_0})$. In this letter we will focus on the
inverted spectrum of neutrinos since only in this case the
predictions coming from leptogenesis depend crucially on the
low-energy phases in the PMNS matrix. The $\Omega$ matrix takes then
the well-known form corresponding to the Type I seesaw with two
right-handed neutrinos~\cite{Ibarra:2003up}:
\begin{equation}
\label{Omega}
\Omega^{\rm IH}=\left(
\begin{array}{cc}
    \pm \sqrt{1-{\omega}^2 } & - {\omega}  \\
  \, {\omega}  & \pm \, \sqrt{1-{\omega}^2 }\\
0&0
\end{array}
\right),
\end{equation}
where $\omega$ is a complex parameter. For the PMNS matrix $U$, we
adopt the usual parametrization~\cite{PDBook}
\begin{equation}
\label{Umatrix}
U=\left( \begin{array}{ccc}
c_{12}\,c_{13} & s_{12}\,c_{13} & s_{13}\,{\rm e}^{-{\rm i}\,\delta} \\
-s_{12}\,c_{23}-c_{12}\,s_{23}\,s_{13}\,{\rm e}^{{\rm i}\,\delta} &
c_{12}\,c_{23}-s_{12}\,s_{23}\,s_{13}\,{\rm e}^{{\rm i}\,\delta} & s_{23}\,c_{13} \\
s_{12}\,s_{23}-c_{12}\,c_{23}\,s_{13}\,{\rm e}^{{\rm i}\,\delta}
& -c_{12}\,s_{23}-s_{12}\,c_{23}\,s_{13}\,{\rm e}^{{\rm i}\,\delta}  &
c_{23}\,c_{13}
\end{array}\right)
\times {\rm diag}(1, {\rm e}^{{\rm i}\,{\Phi/ 2}}, 1)
\, ,
\end{equation}
where $s_{ij}\equiv \sin\theta_{ij}$, $c_{ij}\equiv\cos\theta_{ij}$, $\delta$
is the Dirac $C\!P$-violating phase, $\Phi$ is the Majorana $C\!P$-violating phase.

In order to complete our discussion, we present the relevant
interactions for leptogenesis:
\begin{equation}
V_{\nu} =  h_{\alpha 1} \ \ell_{\alpha}^T \ {\rm i} \sigma_2 \ C \ \rho_3 \ H \ + \
 h_{\alpha 2} \ \ell_{\alpha}^T \ {\rm i} \sigma_2 \  C \ \rho_0 \ H \ + \
 M_{\rho_3} \ \text{Tr} \ \rho_3^T \ C \ \rho_3
\ + \ \frac{1}{2} \ M_{\rho_0} \ \rho_0^T \ C \ \rho_0 \ + \ \text{h.c.}
\end{equation}
and
\begin{eqnarray} \label{scattint}
{\cal{L}}_{\rm kin} & = & {\rm i}\,\text{Tr} \ \bar{\rho}_3 \ \gamma^\mu \ D_\mu \ \rho_3 \ +
{\rm i}\, \bar{\rho}_0 \ \gamma^\mu \ \partial_\mu \ \rho_0
\end{eqnarray}
where $D_\mu \ \rho_3 = \partial_\mu \rho_3 \ + \ {\rm i} g_2 \ [W_\mu, \rho_3]$,
\begin{equation}
\rho_3 = \frac{1}{2} \left( \begin{array} {cc}
 T^0  &  \sqrt{2} \ T^+ \\
 \sqrt{2} \ T^-  & - T^0
\end{array} \right), \qquad \text{and} \qquad
W_\mu = \frac{1}{2} \left( \begin{array} {cc}
 W_\mu^3  &  \sqrt{2} \ W^+_\mu \\
 \sqrt{2} \ W^-_\mu  & - W_\mu^3
\end{array} \right).
\end{equation}

As pointed out in~\cite{Perez:2008ry}, in this model the lightest
field responsible for the seesaw mechanism is the triplet $\rho_3$
and the mass splitting with the singlet $\rho_0$ is very large,
$M_{\rho_0}/M_{\rho_3} > 40$. In~\cite{Blanchet:2008cj} leptogenesis
in this theory was studied in detail. Our findings were that the
$C\!P$ asymmetry is given only by the vertex correction, and that
the case of inverted spectrum for light neutrinos is only marginally
allowed. In the next section, we will focus on this spectrum and
show explicitly that the existence and size of the region of
successful leptogenesis depends crucially on the presence of
low-energy $C\!P$ violation.

\section{Low-energy $C\!P$-violating phases and leptogenesis}
Without assuming any particular texture in the Dirac mass matrix, it
is well-known that the low-energy phases in the PMNS matrix can only
play a role when one considers \emph{flavored} leptogenesis, for
$M_{\rho_3} < 5\times 10^{11}~{\rm GeV}$~\cite{Blanchet:2006ch}. In
the unflavored regime, for $M_{\rho_3}> 5\times 10^{11}~{\rm GeV}$,
the calculation is completely independent of the PMNS phases,
because it involves only the combination $(h^{\dagger}h)_{ij}$,
where the PMNS matrix cancels out. This can be easily verified using
the parametrization given in Eq.~(\ref{ci}). It was shown
in~\cite{Blanchet:2008cj} that the case of inverted hierarchy in
Adjoint $SU(5)$ is only marginally allowed by successful flavored
leptogenesis and that there is no region allowed in the unflavored
regime. It is worth noting that this is in contrast to the 2RH
neutrino model, where a region in the unflavored regime survives. In
the case of normal hierarchy for neutrinos, there in a large region
allowed in the unflavored regime~\cite{Blanchet:2008cj}, so the PMNS
phases only change marginally the constraints. Therefore, as already
mentioned previously, this letter will be exclusively devoted to the
case of inverted hierarchy.

Let us recall the two fundamental quantities for leptogenesis, the
decay parameter and the $C\!P$ asymmetry parameter. In terms of the
orthogonal parametrization given in Eq.~(\ref{ci}), the decay
parameters for the inverted hierarchy are given by
\begin{eqnarray}
\label{param1} K_{\alpha}&=&{m_{1}|U_{\alpha 1}|^2 \over
m_{\star}}|1-\omega^2 | + {m_2|U_{\alpha 2}|^2 \over
m_{\star}}|\omega^2 | \pm {2\sqrt{m_1 m_2} \over m_{\star}} {\rm Re}
\left(U_{\alpha1}U_{\alpha2}^{\star} \sqrt{1-\omega^2 }\omega
^{\star}\right),\\\label{param2} K&=&{m_{1}\over
m_{\star}}|1-\omega^2 | + {m_2 \over m_{\star}}|\omega^2 |,
\end{eqnarray}
where $m_{\star}\simeq 1.08\times 10^{-3}\,{\rm eV}$ is the
equilibrium neutrino mass. As for the flavored $C\!P$ asymmetries
they can be written as
\begin{eqnarray}\label{param3}
\varepsilon_{\rho_3,\alpha}&\simeq& -{1\over 8 \pi v_0^2}{M_{\rho_3} \over m_1 |1-\omega ^2|
+ m_2 |\omega |^2}\left[ \left(m_2^2 |U_{\alpha 2}|^2 - m_1^2 |U_{\alpha 1}|^2\right)
{\rm Im} (\omega ^2)\right. \nonumber\\
&&\pm \sqrt{m_1 m_2}(m_2 + m_1)\,{\rm Re} (U_{\alpha 1}^{\star}
U_{\alpha 2})\,
{\rm Im} \left(\omega \sqrt{1-\omega ^2}\right)\nonumber \\
&&\left.\pm \sqrt{m_1 m_2}(m_2-m_1)\,{\rm Im}( U_{\alpha 1}^{\star}
U_{\alpha 2})\, {\rm Re} \left(\omega \sqrt{1-\omega
^2}\right)\right]. \label{flavCP}
\end{eqnarray}
It can be noticed that the decay parameter and the $C\!P$ asymmetry
(up to a factor 3) are exactly the same as in the 2RH neutrino
model.

As is well known in the 2RH neutrino case, the total $C\!P$
asymmetry
$\varepsilon_{\rho_3}=\sum_{\alpha}\varepsilon_{\rho_3,\alpha}$ in
the case of inverted hierarchy is suppressed by a factor $\Delta
m^2_{\rm sol}/\Delta m^2_{\rm atm}$ compared to normal
hierarchy~\cite{Petcov:2005jh,Blanchet:2006dq}. Additionally, the
total washout parameter $K=\sum_{\alpha}K_{\alpha}$ is typically
larger. Therefore, the lower bounds on the scale of leptogenesis in
the case of inverted hierarchy are much more restrictive than in the
case of normal hierarchy. In flavored leptogenesis, however, there
exists the possibility of having both a smaller flavored decay
parameter, and a flavored $C\!P$ asymmetry that is larger than the
total one, leading to a dramatic decrease of the lowest bound from
$2\times 10^{13}$~GeV to $10^{10}$~GeV~\cite{Blanchet:2008pw}. The
point is that the presence of low-energy phases is essential to keep
the $C\!P$ asymmetry in flavor $\alpha$ large, i.e.
$\varepsilon_{\rho_3,\alpha}\gg \varepsilon_{\rho_3}$, and obtain at
the same time a cancellation in the flavored decay parameter,
leading to $K_{\alpha}\ll K$. This is precisely what we shall obtain
below.

Before turning to the actual evaluation of the role of the
low-energy phases, we would like to point out that the allowed
region obtained in~\cite{Blanchet:2008cj} in the case of inverted
hierarchy does not actually satisfy the condition of validity of the
flavored Boltzmann equations specified in~\cite{Blanchet:2006ch}.
The reason is that the presence of such an allowed region relies
crucially on the fact that the washout is largely reduced by flavor
effects. In this particular situation, the reliability of classical
Boltzmann equation is dubious. However, since a precise description
of leptogenesis in the transition between the flavored and the
unflavored regime, probably relying on a density matrix equation, is
still missing, we shall simply assume here the validity of flavored
Boltzmann equations below $5\times 10^{11}~{\rm GeV}$. This will
enable us to show explicitly that, in the well-motivated model we
consider, flavored leptogenesis does depend on the PMNS phases.

For the details of the leptogenesis computation and all definitions,
we refer the reader to~\cite{Blanchet:2008cj}. We just recall here
that the border of the region of successful leptogenesis is defined
by
\begin{equation}
\label{etaBobs} \eta_B= 5.75\times 10^{-10} \, ,
\end{equation}
corresponding to the lower value allowed by WMAP5 at
$3\sigma$~\cite{Komatsu:2008hk}. Note moreover that we shall use a
slightly different numerical factor compared
to~\cite{Blanchet:2008cj} for the conversion of a $B-L$ asymmetry
into a $B$ asymmetry, 12/37~\cite{Harvey:1990qw} instead of
29/78~\cite{Khlebnikov:1988sr}, assuming that sphalerons remain in
equilibrium until after the electroweak phase transition. This
modifies the relation between the baryon-to-photon ratio $\eta_B$
and the relevant parameter in the Boltzmann equation
$N_{\Delta_{\alpha}}^{\rm f}$ as follows
\begin{equation}\label{etaB}
\eta_B\simeq 3\times0.88\times 10^{-2} \sum_{\alpha}
N_{\Delta_{\alpha}}^{\rm f}.
\end{equation}

Now, in order to understand the role of the low-energy
$C\!P$-violating phases let us define several scenarios:

\begin{itemize}

\item {\underline{\textit{Benchmark I}}}. First, we show in Fig.~\ref{fig:NoPhase}
the result of the flavored computation taking the $C\!P$-conserving
values $\delta=0$ and $\Phi=0$.
\begin{figure}
\begin{center}
\includegraphics[width=0.4\textwidth, angle=-90]{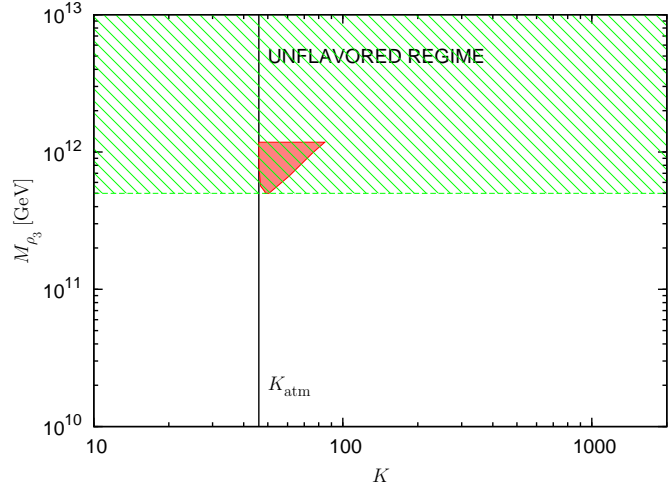}
\caption{The allowed range for the mass of $\rho_3$ vs. the decay
parameter $K$ is the part of the pink region below the hatched area
defining the unflavored regime. Case of inverted hierarchy with
$\Phi=0$, $\delta=0$ and $\sin \theta_{13}=0.2$.}
\label{fig:NoPhase}
\end{center}
\end{figure}
Note that, in order to see more clearly the transition with the next
scenarios, we extended the flavored calculation (pink region) to
$M_{\rho_3}=10^{12}$~GeV. One notices from Fig.~\ref{fig:NoPhase}
that the pink region falls only in the unflavored regime, where our
computation is not valid. Therefore, there is no allowed region.
Changing the value of $\sin \theta_{13}$ does not change this
conclusion. Therefore, low-energy $C\!P$ violation is necessary to
extend the region below the line separating the unflavored from the
flavored regime.

\item {\underline{\textit{Benchmark II}}}. Now we turn on the
Majorana phase $\Phi$, keeping $\delta=0$. The result is shown in
Fig.~\ref{fig:Majo} for two choices of $\sin \theta_{13}$, 0.2 (left
panel) and 0 (right panel).
\begin{figure}
\begin{center}
\includegraphics[width=0.33\textwidth, angle=-90]{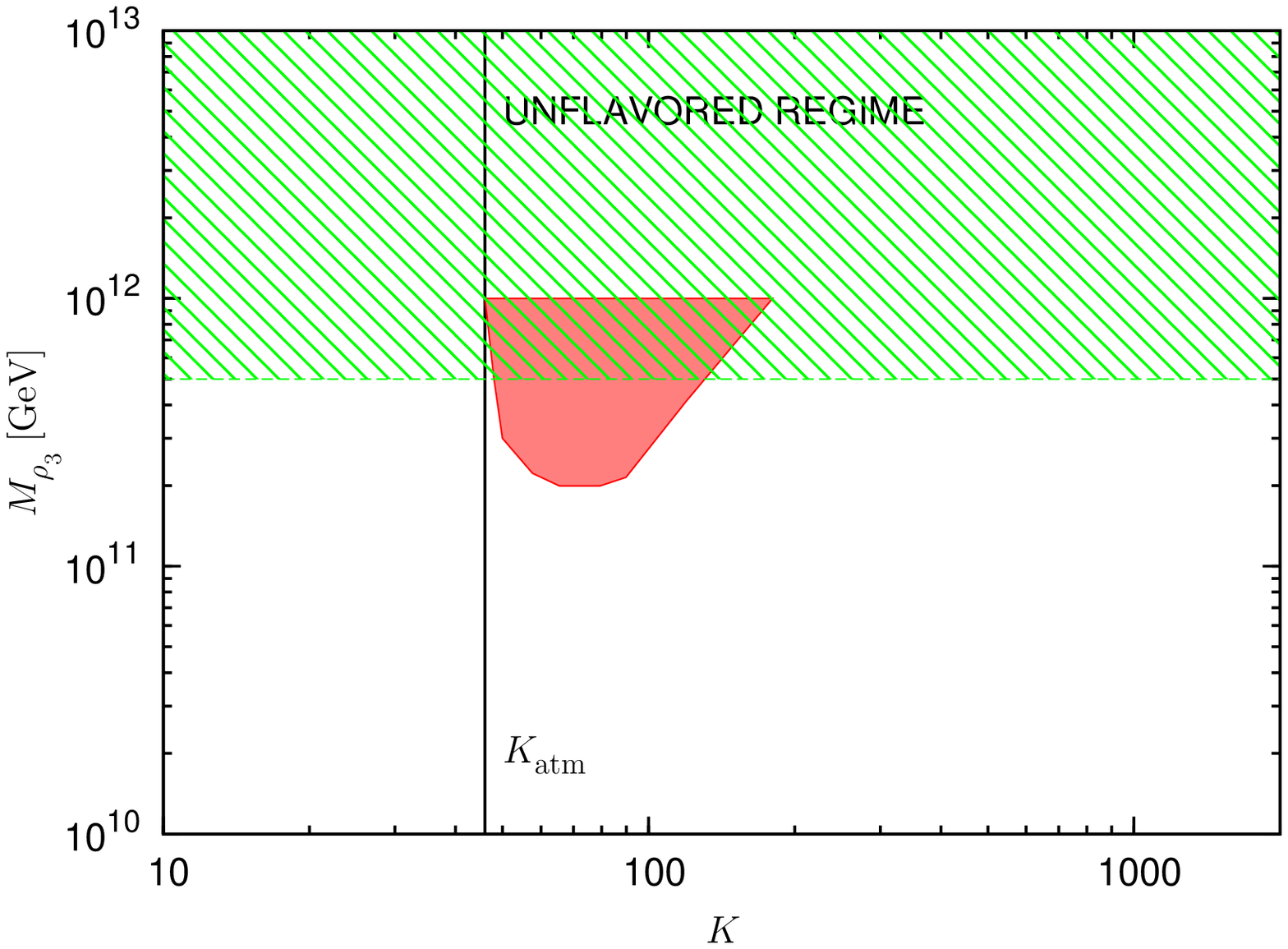}
\hspace{0.3cm}
\includegraphics[width=0.33\textwidth, angle=-90]{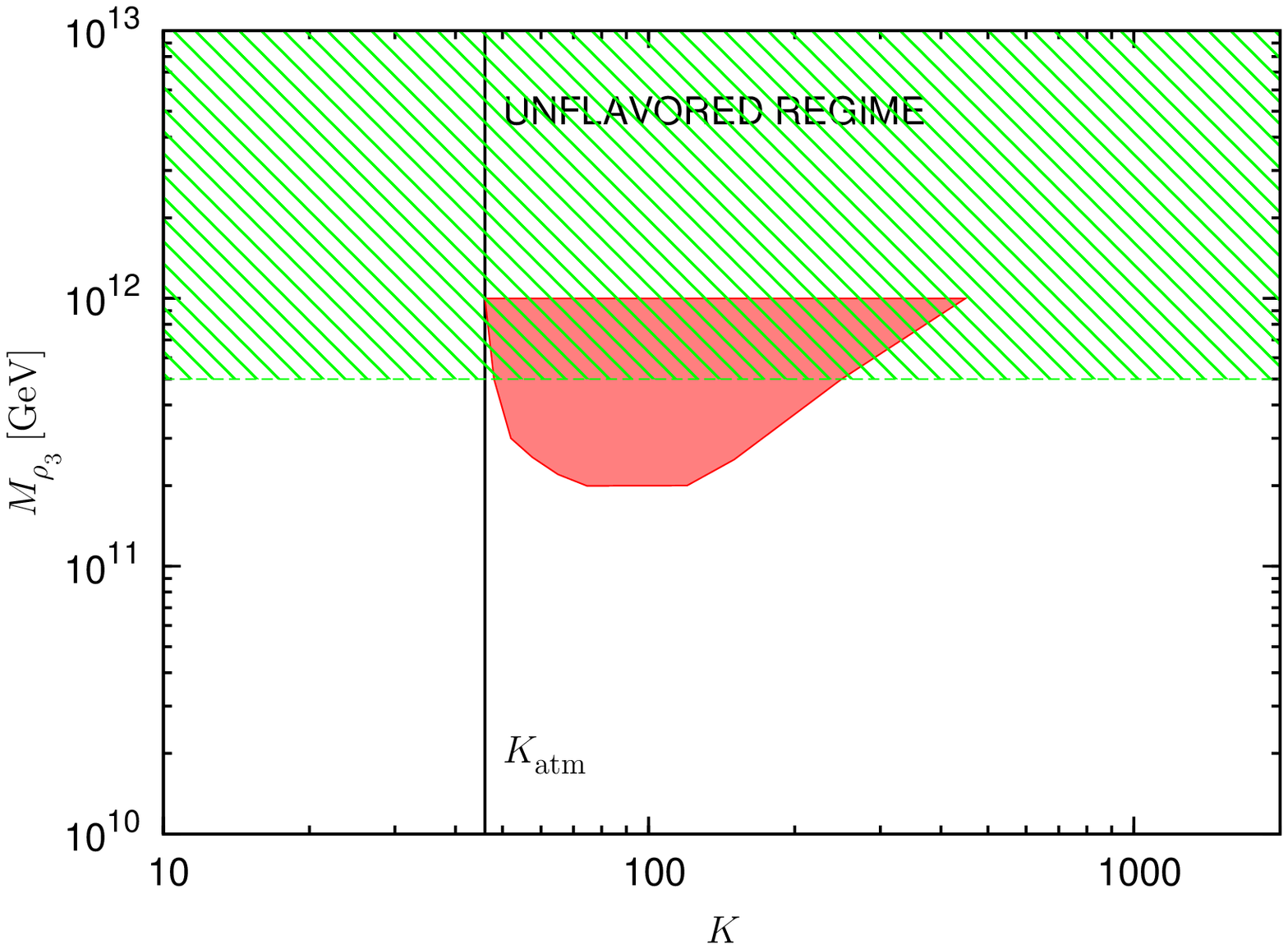}
\caption{Same as Fig.~\ref{fig:NoPhase} but for non-zero Majorana
phase $\Phi$, $\delta=0$, and two values of $\sin \theta_{13}$: 0.2
(left) and 0 (right).} \label{fig:Majo}
\end{center}
\end{figure}
It can be seen that the pink region extends now considerably into
the flavored regime. One actually recovers here the lowest bound
found in~\cite{Blanchet:2008cj} $M_{\rho_3}>2\times 10^{11}$~GeV,
for both values of $\sin \theta_{13}$. Note also that the allowed
region is actually larger for $\sin \theta_{13}=0$.

\item {\underline{\textit{Benchmark III}}}. Finally, we set the Majorana phase to zero,
and turn on the Dirac phase.
The result is shown in Fig.~\ref{fig:Dir} again for two values of
$\sin \theta_{13}$, 0.2 (left panel) and 0.1 (right panel).
\begin{figure}
\begin{center}
\includegraphics[width=0.33\textwidth, angle=-90]{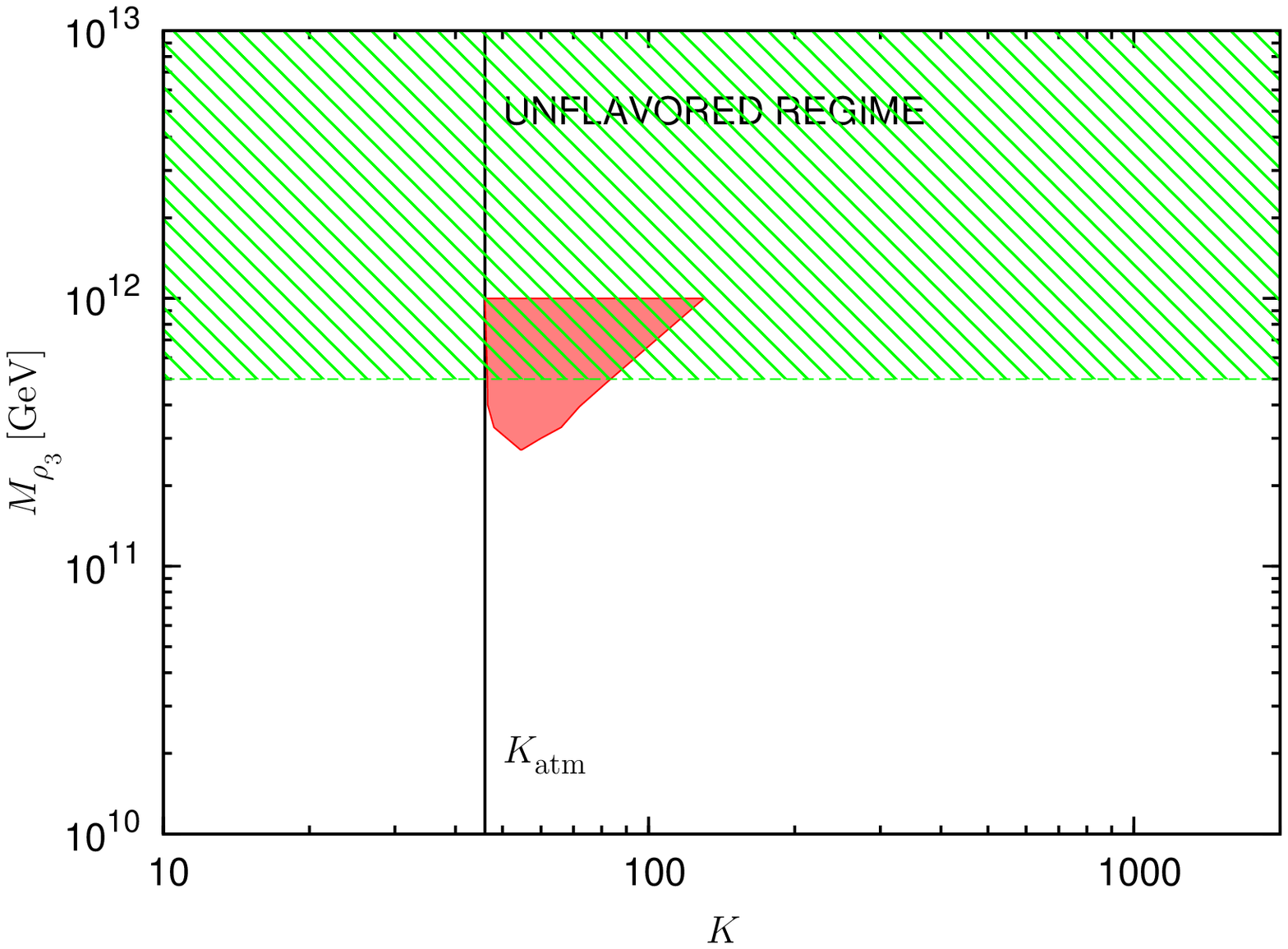}
\hspace{0.3cm}
\includegraphics[width=0.33\textwidth, angle=-90]{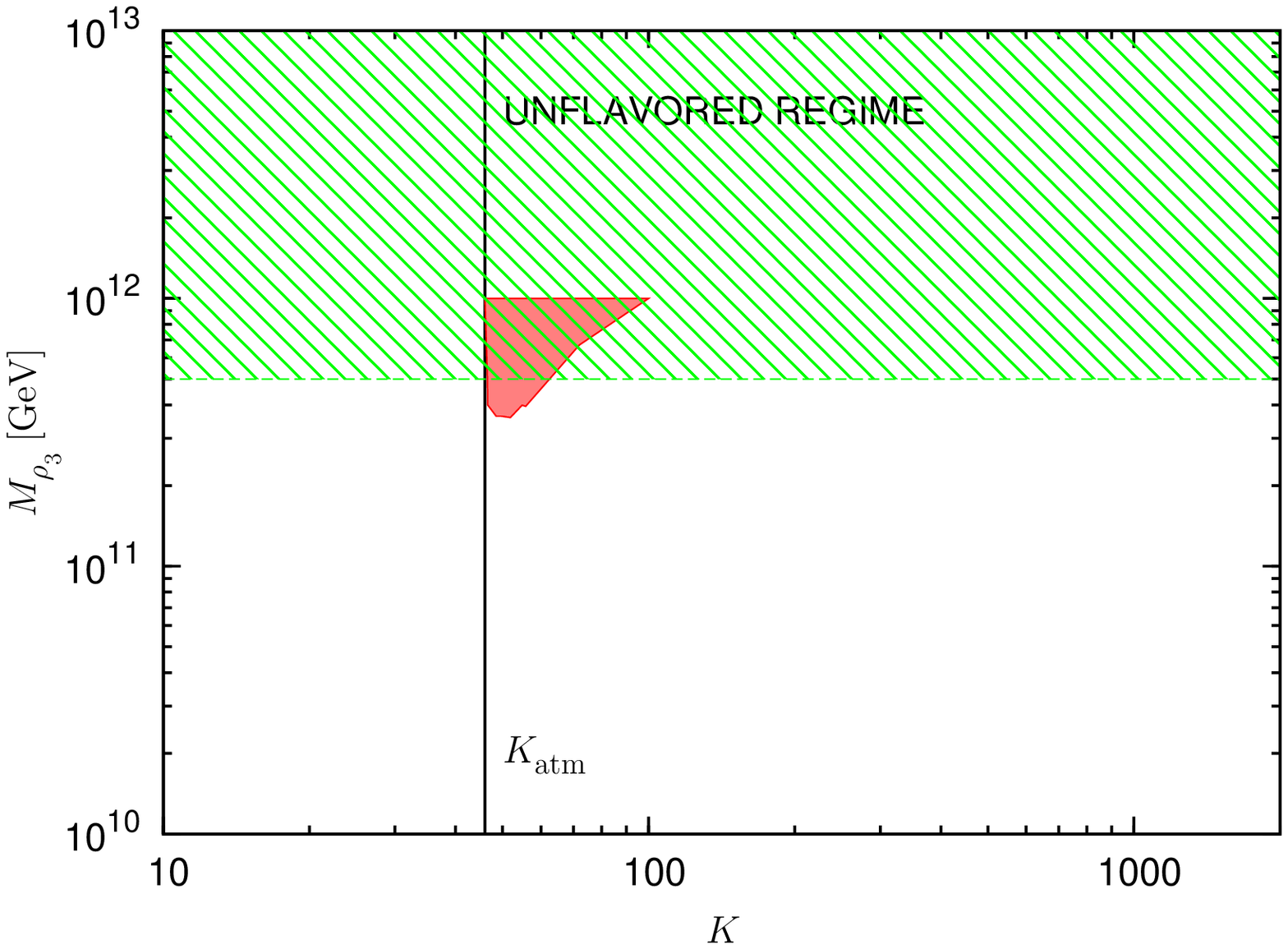}
\caption{Same as Fig.~\ref{fig:NoPhase} but for non-zero Dirac phase
$\delta$, $\Phi=0$, and two values of $\sin \theta_{13}$: 0.2 (left)
and 0.1 (right).} \label{fig:Dir}
\end{center}
\end{figure}
An allowed region opens up below the hatched area also in this case,
even though it remains smaller than in the case of non-zero Majorana
phase, even for the maximal allowed value $\sin \theta_{13}=0.2$.
\end{itemize}

\begin{figure}
\begin{center}
\includegraphics[width=0.45\textwidth, angle=-90]{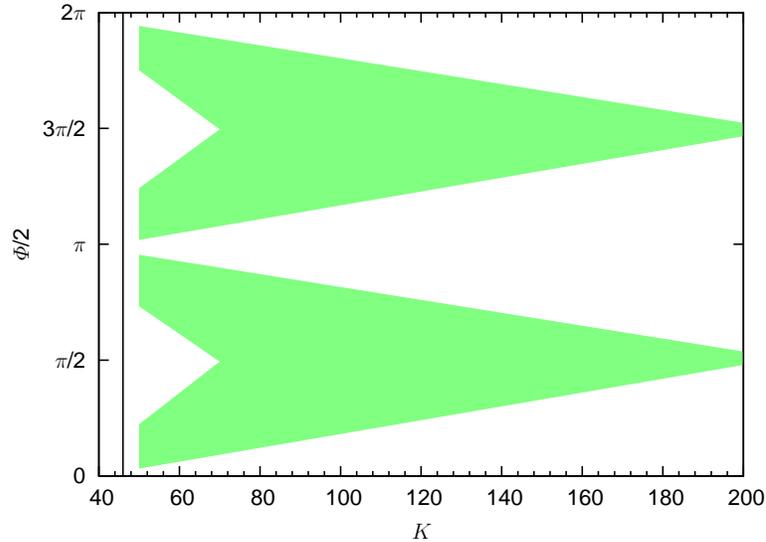}
\caption{Allowed region in the space ($K,\Phi$), fixing
$M_{\rho_3}=3.5\times 10^{11}$~GeV and $\sin \theta_{13}=0$.}
\label{fig:KvsPhi}
\end{center}
\end{figure}

\begin{figure}
\begin{center}
\includegraphics[width=0.45\textwidth, angle=-90]{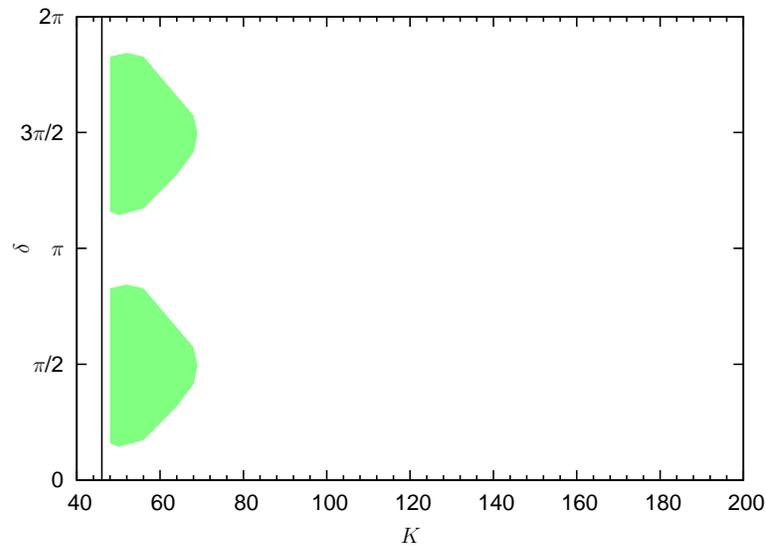}
\caption{Allowed region in the space ($K,\delta$), fixing
$M_{\rho_3}=3.5\times 10^{11}$~GeV, $\Phi=0$ and $\sin
\theta_{13}=0.2$.} \label{fig:KvsDelta}
\end{center}
\end{figure}

In order to illustrate the effect of the PMNS phases in a more
precise way, it is useful to show the allowed regions in the planes
$(K,\Phi)$ and $(K,\delta)$, fixing the mass scale to the value
$M_{\rho_3}=3.5\times 10^{11}$~GeV. The case with only the Majorana
phase is depicted in Fig.~\ref{fig:KvsPhi}, and the case with only
the Dirac phase in Fig.~\ref{fig:KvsDelta}. It is not surprising
that the region is much larger in the case of non-zero Majorana
phase, and that values up to $K=200$ are possible, in agreement with
the right panel of Fig.~\ref{fig:Majo} where the case $\sin
\theta_{13}=0$ is shown. In the case of non-zero Dirac phase with
maximal allowed $\sin \theta_{13}$, only regions around $\delta
=\pi/2,~3\pi/2$ are possible, and at low values of the decay
parameter, $K\lesssim 70$.

Before concluding, it is worth pointing out that our results for the
shape (not the magnitude!) of the lower bounds for different values
of the PMNS phases would be very similar in a 2RH neutrino model.
The main difference comes from the fact that we have here an overall
suppression of the final asymmetry by a factor roughly
$3^{1.2}$~\cite{Blanchet:2008cj}\footnote{The reduction of the
$C\!P$ asymmetry by a factor 3 in this model is compensated by the 3
degrees of freedom of the fermionic triplet as shown in
Eq.~(\ref{etaB}), but the washout is still a factor 3 larger,
leading to the suppression by a factor $3^{1.2}$ in the strong
washout.}, as well as an additional reduction of the efficiency
factor due to the gauge interactions of the fermionic triplet.
Altogether, we have a reduction of the final asymmetry by about one
order of magnitude, leading to the interesting possibility of a
lower bound on $M_{\rho_3}$ very close to the limit of the flavored
regime at $5\times 10^{11}$~GeV, unlike the 2RH neutrino case.
Therefore, the sensitivity to low-energy $C\!P$ violation, as we
have shown, is all the more interesting.

\section{Proton Decay, Unification and Leptogenesis in Adjoint SU(5)}
Leptogenesis is usually considered an elegant but hardly testable
mechanism to explain the matter-antimatter asymmetry of the Universe
due to the very high scale it involves. In this section we present
an interesting scenario which could provide a hint for leptogenesis.

\begin{figure}
\begin{center}
\includegraphics[width=0.7\textwidth, angle=0]{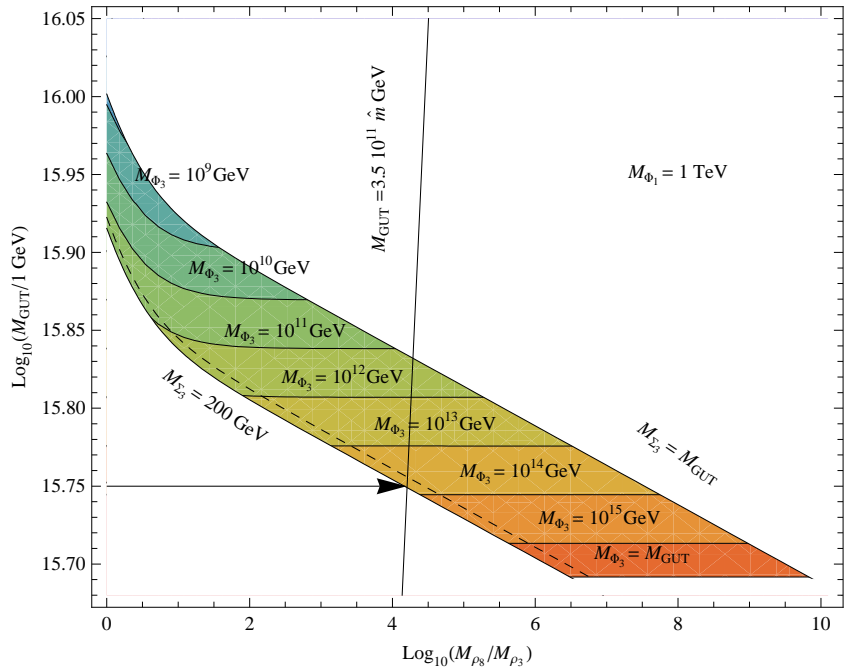}
\caption{Constraints coming from unification of gauge couplings when
$M_{\Phi_1}=1 $ TeV. The field $\Phi_3 \sim (3,3,-1/3)$ mediates
proton decay and its lower bound is $M_{\Phi_3} > 10^{12}$ GeV. The
almost vertical line defines the constraint from leptogenesis in the
inverted spectrum for neutrinos.} \label{unification}
\end{center}
\end{figure}

We start with the optimistic assumption that two states in the
theory of Adjoint $SU(5)$ can be discovered at the LHC, namely the
color scalar octet, $\Phi_1 \sim (8,2,1/2)\subset
\bf{45_H}$~\cite{Perez:2008ib} and the scalar $SU(2)_L$ triplet
$\Sigma_3 \sim (1,3,0)\subset \bf{24_H}$. Then, as we have seen in
the last section, leptogenesis for the inverted scheme of light
neutrinos occurs for a very restricted mass range around
$M_{\rho_3}=3.5\times 10^{11}$~GeV. So we can take the constraint
from leptogenesis as a line in parameter space, instead of an
extended region as in the case of normal
hierarchy~\cite{Blanchet:2008cj}. Fixing the mass
$M_{\Phi_1}=1$~TeV, we show in Fig.~\ref{unification} how the
constraints from gauge coupling unification and leptogenesis when
$M_{\Sigma_3}=200$~GeV allow to pin down the GUT scale. We find
$M_{\rm GUT}=5.62 \times 10^{15}$~GeV, from which the following
proton lifetimes can be derived~\cite{Perez:2008ry}: $\tau (p \to
e^+ \pi^0)= 3.2 \times 10^{34}$~years, $\tau (p \to K^+ \bar{\nu})=
2.5 \times 10^{36}$~years and $\tau (p \to \pi^+ \bar{\nu})=8.1
\times 10^{34}$~years. These lifetimes are potentially observable at
future experiments~\cite{Autiero:2007zj}.

It is very interesting that Adjoint $SU(5)$ provides a framework
which allows to relate so different concepts as proton decay,
leptogenesis and gauge couplings unification. We find that, assuming
an inverted spectrum for neutrinos and unification of gauge
couplings at a certain scale, the observation of proton decay in the
predicted range can provide an indication for the high scale of
leptogenesis.

\section{Conclusion}
We have investigated the effect of the low-energy $C\!P$-violating
phases on leptogenesis in the context of Adjoint $SU(5)$. In this
model neutrino masses are generated through the Type~I and Type~III
seesaw mechanisms, and the lepton asymmetry is generated by the
fermionic triplet responsible for the Type III seesaw. We have found
that, for an inverted spectrum of neutrinos, leptogenesis is
disfavored without low-energy $C\!P$-violating phases assuming the
validity of flavored Boltzmann equations below the usual scale of
$5\times 10^{11}$~GeV. Therefore, if one discovers that the spectrum
of light neutrinos is inverted, Adjoint $SU(5)$ offers a nice
example of a theory where leptogenesis is very sensitive to
low-energy $C\!P$ violation. We have shown several numerical
examples in order to understand the precise impact of each one of
the PMNS phases. Finally, we have presented a scenario where the
observation of new states at the LHC, together with unification of
gauge couplings and proton decay in the predicted range, can provide
an indication for the high scale of leptogenesis.
\begin{acknowledgments}
We would like to thank Pasquale Di Bari for discussions. The work of
P. F. P. was supported in part by the U.S. Department of Energy
contract No. DE-FG02-08ER41531 and in part by the Wisconsin Alumni
Research Foundation.
\end{acknowledgments}


\begin{thebibliography}{42}
\expandafter\ifx\csname
natexlab\endcsname\relax\def\natexlab#1{#1}\fi
\expandafter\ifx\csname bibnamefont\endcsname\relax
  \def\bibnamefont#1{#1}\fi
\expandafter\ifx\csname bibfnamefont\endcsname\relax
  \def\bibfnamefont#1{#1}\fi
\expandafter\ifx\csname citenamefont\endcsname\relax
  \def\citenamefont#1{#1}\fi
\expandafter\ifx\csname url\endcsname\relax
  \def\url#1{\texttt{#1}}\fi
\expandafter\ifx\csname urlprefix\endcsname\relax\def\urlprefix{URL
}\fi \providecommand{\bibinfo}[2]{#2}
\providecommand{\eprint}[2][]{\url{#2}}

\bibitem[{\citenamefont{Fukugita and Yanagida}(1986)}]{Fukugita:1986hr}
\bibinfo{author}{\bibfnamefont{M.}~\bibnamefont{Fukugita}} \bibnamefont{and}
  \bibinfo{author}{\bibfnamefont{T.}~\bibnamefont{Yanagida}},
  \bibinfo{journal}{Phys. Lett.} \textbf{\bibinfo{volume}{B174}},
  \bibinfo{pages}{45} (\bibinfo{year}{1986}).

\bibitem[{\citenamefont{Minkowski}(1977)}]{Minkowski:1977sc}
\bibinfo{author}{\bibfnamefont{P.}~\bibnamefont{Minkowski}},
  \bibinfo{journal}{Phys. Lett.} \textbf{\bibinfo{volume}{B67}},
  \bibinfo{pages}{421} (\bibinfo{year}{1977}).

\bibitem[{\citenamefont{Yanagida}(1979)}]{Yanagida}
\bibinfo{author}{\bibfnamefont{T.}~\bibnamefont{Yanagida}}, in
  \emph{\bibinfo{booktitle}{Workshop on Unified Theories, KEK Report 79-18}}
  (\bibinfo{year}{1979}), p.~\bibinfo{pages}{95}.

\bibitem[{\citenamefont{Gell-Mann et~al.}(1979)\citenamefont{Gell-Mann, Ramond,
  and Slansky}}]{Gell-Mann}
\bibinfo{author}{\bibfnamefont{M.}~\bibnamefont{Gell-Mann}},
  \bibinfo{author}{\bibfnamefont{P.}~\bibnamefont{Ramond}}, \bibnamefont{and}
  \bibinfo{author}{\bibfnamefont{R.}~\bibnamefont{Slansky}},
  \emph{\bibinfo{title}{Supergravity}} (\bibinfo{publisher}{Amsterdam: North
  Holland}, \bibinfo{year}{1979}), p. \bibinfo{pages}{315}.

\bibitem[{\citenamefont{Glashow}(1980)}]{Glashow}
\bibinfo{author}{\bibfnamefont{S.~L.} \bibnamefont{Glashow}},
  \emph{\bibinfo{title}{1979 Cargese Summer Institute on Quarks and Leptons}}
  (\bibinfo{publisher}{New York: Plenum}, \bibinfo{year}{1980}), p.
  \bibinfo{pages}{687}.

\bibitem[{\citenamefont{Barbieri et~al.}(1980)\citenamefont{Barbieri,
  Nanopoulos, Morchio, and Strocchi}}]{Barbieri:1979ag}
\bibinfo{author}{\bibfnamefont{R.}~\bibnamefont{Barbieri}},
  \bibinfo{author}{\bibfnamefont{D.~V.} \bibnamefont{Nanopoulos}},
  \bibinfo{author}{\bibfnamefont{G.}~\bibnamefont{Morchio}}, \bibnamefont{and}
  \bibinfo{author}{\bibfnamefont{F.}~\bibnamefont{Strocchi}},
  \bibinfo{journal}{Phys. Lett.} \textbf{\bibinfo{volume}{B90}},
  \bibinfo{pages}{91} (\bibinfo{year}{1980}).

\bibitem[{\citenamefont{Mohapatra and Senjanovic}(1981)}]{Mohapatra:1980yp}
\bibinfo{author}{\bibfnamefont{R.~N.} \bibnamefont{Mohapatra}}
  \bibnamefont{and}
  \bibinfo{author}{\bibfnamefont{G.}~\bibnamefont{Senjanovic}},
  \bibinfo{journal}{Phys. Rev.} \textbf{\bibinfo{volume}{D23}},
  \bibinfo{pages}{165} (\bibinfo{year}{1981}).

\bibitem[{\citenamefont{Branco et~al.}(2001)\citenamefont{Branco, Morozumi,
  Nobre, and Rebelo}}]{Branco:2001pq}
\bibinfo{author}{\bibfnamefont{G.~C.} \bibnamefont{Branco}},
  \bibinfo{author}{\bibfnamefont{T.}~\bibnamefont{Morozumi}},
  \bibinfo{author}{\bibfnamefont{B.~M.} \bibnamefont{Nobre}}, \bibnamefont{and}
  \bibinfo{author}{\bibfnamefont{M.~N.} \bibnamefont{Rebelo}},
  \bibinfo{journal}{Nucl. Phys.} \textbf{\bibinfo{volume}{B617}},
  \bibinfo{pages}{475} (\bibinfo{year}{2001}), \eprint{hep-ph/0107164}.

\bibitem[{\citenamefont{Frampton et~al.}(2002)\citenamefont{Frampton, Glashow,
  and Yanagida}}]{Frampton:2002qc}
\bibinfo{author}{\bibfnamefont{P.~H.} \bibnamefont{Frampton}},
  \bibinfo{author}{\bibfnamefont{S.~L.} \bibnamefont{Glashow}},
  \bibnamefont{and} \bibinfo{author}{\bibfnamefont{T.}~\bibnamefont{Yanagida}},
  \bibinfo{journal}{Phys. Lett.} \textbf{\bibinfo{volume}{B548}},
  \bibinfo{pages}{119} (\bibinfo{year}{2002}), \eprint{hep-ph/0208157}.

\bibitem[{\citenamefont{Davidson et~al.}(2008)\citenamefont{Davidson, Nardi,
  and Nir}}]{Davidson:2008bu}
\bibinfo{author}{\bibfnamefont{S.}~\bibnamefont{Davidson}},
  \bibinfo{author}{\bibfnamefont{E.}~\bibnamefont{Nardi}}, \bibnamefont{and}
  \bibinfo{author}{\bibfnamefont{Y.}~\bibnamefont{Nir}} (\bibinfo{year}{2008}),
  \eprint{0802.2962}.

\bibitem[{\citenamefont{Abada et~al.}(2006)\citenamefont{Abada, Davidson,
  Josse-Michaux, Losada, and Riotto}}]{Abada:2006fw}
\bibinfo{author}{\bibfnamefont{A.}~\bibnamefont{Abada}},
  \bibinfo{author}{\bibfnamefont{S.}~\bibnamefont{Davidson}},
  \bibinfo{author}{\bibfnamefont{F.-X.} \bibnamefont{Josse-Michaux}},
  \bibinfo{author}{\bibfnamefont{M.}~\bibnamefont{Losada}}, \bibnamefont{and}
  \bibinfo{author}{\bibfnamefont{A.}~\bibnamefont{Riotto}},
  \bibinfo{journal}{JCAP} \textbf{\bibinfo{volume}{0604}}, \bibinfo{pages}{004}
  (\bibinfo{year}{2006}), \eprint{hep-ph/0601083}.

\bibitem[{\citenamefont{Nardi et~al.}(2006)\citenamefont{Nardi, Nir, Roulet,
  and Racker}}]{Nardi:2006fx}
\bibinfo{author}{\bibfnamefont{E.}~\bibnamefont{Nardi}},
  \bibinfo{author}{\bibfnamefont{Y.}~\bibnamefont{Nir}},
  \bibinfo{author}{\bibfnamefont{E.}~\bibnamefont{Roulet}}, \bibnamefont{and}
  \bibinfo{author}{\bibfnamefont{J.}~\bibnamefont{Racker}},
  \bibinfo{journal}{JHEP} \textbf{\bibinfo{volume}{01}}, \bibinfo{pages}{164}
  (\bibinfo{year}{2006}), \eprint{hep-ph/0601084}.

\bibitem[{\citenamefont{Davidson et~al.}(2007)\citenamefont{Davidson, Garayoa,
  Palorini, and Rius}}]{Davidson:2007va}
\bibinfo{author}{\bibfnamefont{S.}~\bibnamefont{Davidson}},
  \bibinfo{author}{\bibfnamefont{J.}~\bibnamefont{Garayoa}},
  \bibinfo{author}{\bibfnamefont{F.}~\bibnamefont{Palorini}}, \bibnamefont{and}
  \bibinfo{author}{\bibfnamefont{N.}~\bibnamefont{Rius}},
  \bibinfo{journal}{Phys. Rev. Lett.} \textbf{\bibinfo{volume}{99}},
  \bibinfo{pages}{161801} (\bibinfo{year}{2007}), \eprint{arXiv:0705.1503
  [hep-ph]}.

\bibitem[{\citenamefont{Blanchet and Di~Bari}(2007)}]{Blanchet:2006be}
\bibinfo{author}{\bibfnamefont{S.}~\bibnamefont{Blanchet}} \bibnamefont{and}
  \bibinfo{author}{\bibfnamefont{P.}~\bibnamefont{Di~Bari}},
  \bibinfo{journal}{JCAP} \textbf{\bibinfo{volume}{0703}}, \bibinfo{pages}{018}
  (\bibinfo{year}{2007}), \eprint{hep-ph/0607330}.

\bibitem[{\citenamefont{Pascoli
  et~al.}(2007{\natexlab{a}})\citenamefont{Pascoli, Petcov, and
  Riotto}}]{Pascoli:2006ie}
\bibinfo{author}{\bibfnamefont{S.}~\bibnamefont{Pascoli}},
  \bibinfo{author}{\bibfnamefont{S.~T.} \bibnamefont{Petcov}},
  \bibnamefont{and} \bibinfo{author}{\bibfnamefont{A.}~\bibnamefont{Riotto}},
  \bibinfo{journal}{Phys. Rev.} \textbf{\bibinfo{volume}{D75}},
  \bibinfo{pages}{083511} (\bibinfo{year}{2007}{\natexlab{a}}),
  \eprint{hep-ph/0609125}.

\bibitem[{\citenamefont{Pascoli
  et~al.}(2007{\natexlab{b}})\citenamefont{Pascoli, Petcov, and
  Riotto}}]{Pascoli:2006ci}
\bibinfo{author}{\bibfnamefont{S.}~\bibnamefont{Pascoli}},
  \bibinfo{author}{\bibfnamefont{S.~T.} \bibnamefont{Petcov}},
  \bibnamefont{and} \bibinfo{author}{\bibfnamefont{A.}~\bibnamefont{Riotto}},
  \bibinfo{journal}{Nucl. Phys.} \textbf{\bibinfo{volume}{B774}},
  \bibinfo{pages}{1} (\bibinfo{year}{2007}{\natexlab{b}}),
  \eprint{hep-ph/0611338}.

\bibitem[{\citenamefont{Branco et~al.}(2007)\citenamefont{Branco,
  Gonzalez~Felipe, and Joaquim}}]{Branco:2006ce}
\bibinfo{author}{\bibfnamefont{G.~C.} \bibnamefont{Branco}},
  \bibinfo{author}{\bibfnamefont{R.}~\bibnamefont{Gonzalez~Felipe}},
  \bibnamefont{and} \bibinfo{author}{\bibfnamefont{F.~R.}
  \bibnamefont{Joaquim}}, \bibinfo{journal}{Phys. Lett.}
  \textbf{\bibinfo{volume}{B645}}, \bibinfo{pages}{432} (\bibinfo{year}{2007}),
  \eprint{hep-ph/0609297}.

\bibitem[{\citenamefont{Anisimov et~al.}(2008)\citenamefont{Anisimov, Blanchet,
  and Di~Bari}}]{Anisimov:2007mw}
\bibinfo{author}{\bibfnamefont{A.}~\bibnamefont{Anisimov}},
  \bibinfo{author}{\bibfnamefont{S.}~\bibnamefont{Blanchet}}, \bibnamefont{and}
  \bibinfo{author}{\bibfnamefont{P.}~\bibnamefont{Di~Bari}},
  \bibinfo{journal}{JCAP} \textbf{\bibinfo{volume}{0804}}, \bibinfo{pages}{033}
  (\bibinfo{year}{2008}), \eprint{0707.3024}.

\bibitem[{\citenamefont{Molinaro et~al.}(2007)\citenamefont{Molinaro, Petcov,
  Shindou, and Takanishi}}]{Molinaro:2007uv}
\bibinfo{author}{\bibfnamefont{E.}~\bibnamefont{Molinaro}},
  \bibinfo{author}{\bibfnamefont{S.~T.} \bibnamefont{Petcov}},
  \bibinfo{author}{\bibfnamefont{T.}~\bibnamefont{Shindou}}, \bibnamefont{and}
  \bibinfo{author}{\bibfnamefont{Y.}~\bibnamefont{Takanishi}}
  (\bibinfo{year}{2007}), \eprint{arXiv:0709.0413 [hep-ph]}.

\bibitem[{\citenamefont{Molinaro and
  Petcov}(2008{\natexlab{a}})}]{Molinaro:2008cw}
\bibinfo{author}{\bibfnamefont{E.}~\bibnamefont{Molinaro}} \bibnamefont{and}
  \bibinfo{author}{\bibfnamefont{S.~T.} \bibnamefont{Petcov}}
  (\bibinfo{year}{2008}{\natexlab{a}}), \eprint{0808.3534}.

\bibitem[{\citenamefont{Ibarra and Ross}(2004)}]{Ibarra:2003up}
\bibinfo{author}{\bibfnamefont{A.}~\bibnamefont{Ibarra}} \bibnamefont{and}
  \bibinfo{author}{\bibfnamefont{G.~G.} \bibnamefont{Ross}},
  \bibinfo{journal}{Phys. Lett.} \textbf{\bibinfo{volume}{B591}},
  \bibinfo{pages}{285} (\bibinfo{year}{2004}), \eprint{hep-ph/0312138}.

\bibitem[{\citenamefont{Chankowski and Turzynski}(2003)}]{Chankowski:2003rr}
\bibinfo{author}{\bibfnamefont{P.~H.} \bibnamefont{Chankowski}}
  \bibnamefont{and}
  \bibinfo{author}{\bibfnamefont{K.}~\bibnamefont{Turzynski}},
  \bibinfo{journal}{Phys. Lett.} \textbf{\bibinfo{volume}{B570}},
  \bibinfo{pages}{198} (\bibinfo{year}{2003}), \eprint{hep-ph/0306059}.

\bibitem[{\citenamefont{Blanchet and Di~Bari}(2008)}]{Blanchet:2008pw}
\bibinfo{author}{\bibfnamefont{S.}~\bibnamefont{Blanchet}} \bibnamefont{and}
  \bibinfo{author}{\bibfnamefont{P.}~\bibnamefont{Di~Bari}}
  (\bibinfo{year}{2008}), \eprint{0807.0743}.

\bibitem[{\citenamefont{Molinaro and
  Petcov}(2008{\natexlab{b}})}]{Molinaro:2008rg}
\bibinfo{author}{\bibfnamefont{E.}~\bibnamefont{Molinaro}} \bibnamefont{and}
  \bibinfo{author}{\bibfnamefont{S.~T.} \bibnamefont{Petcov}}
  (\bibinfo{year}{2008}{\natexlab{b}}), \eprint{0803.4120}.

\bibitem[{\citenamefont{Di~Bari and Riotto}(2008)}]{DiBari:2008mp}
\bibinfo{author}{\bibfnamefont{P.}~\bibnamefont{Di~Bari}} \bibnamefont{and}
  \bibinfo{author}{\bibfnamefont{A.}~\bibnamefont{Riotto}}
  (\bibinfo{year}{2008}), \eprint{0809.2285}.

\bibitem[{\citenamefont{Abada et~al.}(2008)\citenamefont{Abada, Hosteins,
  Josse-Michaux, and Lavignac}}]{Abada:2008gs}
\bibinfo{author}{\bibfnamefont{A.}~\bibnamefont{Abada}},
  \bibinfo{author}{\bibfnamefont{P.}~\bibnamefont{Hosteins}},
  \bibinfo{author}{\bibfnamefont{F.-X.} \bibnamefont{Josse-Michaux}},
  \bibnamefont{and} \bibinfo{author}{\bibfnamefont{S.}~\bibnamefont{Lavignac}}
  (\bibinfo{year}{2008}), \eprint{0808.2058}.

\bibitem[{\citenamefont{Fileviez~P\'erez}(2007)}]{Perez:2007rm}
\bibinfo{author}{\bibfnamefont{P.}~\bibnamefont{Fileviez~P\'erez}},
  \bibinfo{journal}{Phys. Lett.} \textbf{\bibinfo{volume}{B654}},
  \bibinfo{pages}{189} (\bibinfo{year}{2007}), \eprint{hep-ph/0702287}.

\bibitem[{\citenamefont{Foot et~al.}(1989)\citenamefont{Foot, Lew, He, and
  Joshi}}]{Foot:1988aq}
\bibinfo{author}{\bibfnamefont{R.}~\bibnamefont{Foot}},
  \bibinfo{author}{\bibfnamefont{H.}~\bibnamefont{Lew}},
  \bibinfo{author}{\bibfnamefont{X.~G.} \bibnamefont{He}}, \bibnamefont{and}
  \bibinfo{author}{\bibfnamefont{G.~C.} \bibnamefont{Joshi}},
  \bibinfo{journal}{Z. Phys.} \textbf{\bibinfo{volume}{C44}},
  \bibinfo{pages}{441} (\bibinfo{year}{1989}).

\bibitem[{\citenamefont{Fileviez~P\'erez
  et~al.}(2008{\natexlab{a}})\citenamefont{Fileviez~P\'erez, Iminniyaz, and
  Rodrigo}}]{Perez:2008ry}
\bibinfo{author}{\bibfnamefont{P.}~\bibnamefont{Fileviez~P\'erez}},
  \bibinfo{author}{\bibfnamefont{H.}~\bibnamefont{Iminniyaz}},
  \bibnamefont{and} \bibinfo{author}{\bibfnamefont{G.}~\bibnamefont{Rodrigo}},
  \bibinfo{journal}{Phys. Rev.} \textbf{\bibinfo{volume}{D78}},
  \bibinfo{pages}{015013} (\bibinfo{year}{2008}{\natexlab{a}}),
  \eprint{0803.4156}.

\bibitem[{\citenamefont{Hambye et~al.}(2004)\citenamefont{Hambye, Lin, Notari,
  Papucci, and Strumia}}]{Hambye:2003rt}
\bibinfo{author}{\bibfnamefont{T.}~\bibnamefont{Hambye}},
  \bibinfo{author}{\bibfnamefont{Y.}~\bibnamefont{Lin}},
  \bibinfo{author}{\bibfnamefont{A.}~\bibnamefont{Notari}},
  \bibinfo{author}{\bibfnamefont{M.}~\bibnamefont{Papucci}}, \bibnamefont{and}
  \bibinfo{author}{\bibfnamefont{A.}~\bibnamefont{Strumia}},
  \bibinfo{journal}{Nucl. Phys.} \textbf{\bibinfo{volume}{B695}},
  \bibinfo{pages}{169} (\bibinfo{year}{2004}), \eprint{hep-ph/0312203}.

\bibitem[{\citenamefont{Blanchet and Fileviez~P\'erez}(2008)}]{Blanchet:2008cj}
\bibinfo{author}{\bibfnamefont{S.}~\bibnamefont{Blanchet}} \bibnamefont{and}
  \bibinfo{author}{\bibfnamefont{P.}~\bibnamefont{Fileviez~P\'erez}},
  \bibinfo{journal}{JCAP} \textbf{\bibinfo{volume}{0808}}, \bibinfo{pages}{037}
  (\bibinfo{year}{2008}), \eprint{0807.3740}.

\bibitem[{\citenamefont{Fileviez~P\'erez
  et~al.}(2008{\natexlab{b}})\citenamefont{Fileviez~P\'erez, Gavin, McElmurry,
  and Petriello}}]{Perez:2008ib}
\bibinfo{author}{\bibfnamefont{P.}~\bibnamefont{Fileviez~P\'erez}},
  \bibinfo{author}{\bibfnamefont{R.}~\bibnamefont{Gavin}},
  \bibinfo{author}{\bibfnamefont{T.}~\bibnamefont{McElmurry}},
  \bibnamefont{and} \bibinfo{author}{\bibfnamefont{F.}~\bibnamefont{Petriello}}
  (\bibinfo{year}{2008}{\natexlab{b}}), \eprint{0809.2106}.

\bibitem[{\citenamefont{Gonzalez-Garcia and
  Maltoni}(2007)}]{GonzalezGarcia:2007ib}
\bibinfo{author}{\bibfnamefont{M.~C.} \bibnamefont{Gonzalez-Garcia}}
  \bibnamefont{and} \bibinfo{author}{\bibfnamefont{M.}~\bibnamefont{Maltoni}}
  (\bibinfo{year}{2007}), \eprint{arXiv:0704.1800 [hep-ph]}.

\bibitem[{\citenamefont{Casas and Ibarra}(2001)}]{Casas:2001sr}
\bibinfo{author}{\bibfnamefont{J.~A.} \bibnamefont{Casas}} \bibnamefont{and}
  \bibinfo{author}{\bibfnamefont{A.}~\bibnamefont{Ibarra}},
  \bibinfo{journal}{Nucl. Phys.} \textbf{\bibinfo{volume}{B618}},
  \bibinfo{pages}{171} (\bibinfo{year}{2001}), \eprint{hep-ph/0103065}.

\bibitem[{\citenamefont{Yao et~al.}(2006)}]{PDBook}
\bibinfo{author}{\bibfnamefont{W.-M.} \bibnamefont{Yao}} \bibnamefont{et~al.},
  \bibinfo{journal}{Journal of Physics G} \textbf{\bibinfo{volume}{33}},
  \bibinfo{pages}{1+} (\bibinfo{year}{2006}),
  \urlprefix\url{http://pdg.lbl.gov}.

\bibitem[{\citenamefont{Blanchet et~al.}(2007)\citenamefont{Blanchet, Di~Bari,
  and Raffelt}}]{Blanchet:2006ch}
\bibinfo{author}{\bibfnamefont{S.}~\bibnamefont{Blanchet}},
  \bibinfo{author}{\bibfnamefont{P.}~\bibnamefont{Di~Bari}}, \bibnamefont{and}
  \bibinfo{author}{\bibfnamefont{G.~G.} \bibnamefont{Raffelt}},
  \bibinfo{journal}{JCAP} \textbf{\bibinfo{volume}{0703}}, \bibinfo{pages}{012}
  (\bibinfo{year}{2007}), \eprint{hep-ph/0611337}.

\bibitem[{\citenamefont{Petcov et~al.}(2006)\citenamefont{Petcov, Rodejohann,
  Shindou, and Takanishi}}]{Petcov:2005jh}
\bibinfo{author}{\bibfnamefont{S.~T.} \bibnamefont{Petcov}},
  \bibinfo{author}{\bibfnamefont{W.}~\bibnamefont{Rodejohann}},
  \bibinfo{author}{\bibfnamefont{T.}~\bibnamefont{Shindou}}, \bibnamefont{and}
  \bibinfo{author}{\bibfnamefont{Y.}~\bibnamefont{Takanishi}},
  \bibinfo{journal}{Nucl. Phys.} \textbf{\bibinfo{volume}{B739}},
  \bibinfo{pages}{208} (\bibinfo{year}{2006}), \eprint{hep-ph/0510404}.

\bibitem[{\citenamefont{Blanchet and Di~Bari}(2006)}]{Blanchet:2006dq}
\bibinfo{author}{\bibfnamefont{S.}~\bibnamefont{Blanchet}} \bibnamefont{and}
  \bibinfo{author}{\bibfnamefont{P.}~\bibnamefont{Di~Bari}},
  \bibinfo{journal}{JCAP} \textbf{\bibinfo{volume}{0606}}, \bibinfo{pages}{023}
  (\bibinfo{year}{2006}), \eprint{hep-ph/0603107}.

\bibitem[{\citenamefont{Komatsu et~al.}(2008)}]{Komatsu:2008hk}
\bibinfo{author}{\bibfnamefont{E.}~\bibnamefont{Komatsu}} \bibnamefont{et~al.}
  (\bibinfo{collaboration}{WMAP}) (\bibinfo{year}{2008}), \eprint{0803.0547}.

\bibitem[{\citenamefont{Harvey and Turner}(1990)}]{Harvey:1990qw}
\bibinfo{author}{\bibfnamefont{J.~A.} \bibnamefont{Harvey}} \bibnamefont{and}
  \bibinfo{author}{\bibfnamefont{M.~S.} \bibnamefont{Turner}},
  \bibinfo{journal}{Phys. Rev.} \textbf{\bibinfo{volume}{D42}},
  \bibinfo{pages}{3344} (\bibinfo{year}{1990}).

\bibitem[{\citenamefont{Khlebnikov and Shaposhnikov}(1988)}]{Khlebnikov:1988sr}
\bibinfo{author}{\bibfnamefont{S.~Y.} \bibnamefont{Khlebnikov}}
  \bibnamefont{and} \bibinfo{author}{\bibfnamefont{M.~E.}
  \bibnamefont{Shaposhnikov}}, \bibinfo{journal}{Nucl. Phys.}
  \textbf{\bibinfo{volume}{B308}}, \bibinfo{pages}{885} (\bibinfo{year}{1988}).

\bibitem[{\citenamefont{Autiero et~al.}(2007)}]{Autiero:2007zj}
\bibinfo{author}{\bibfnamefont{D.}~\bibnamefont{Autiero}} \bibnamefont{et~al.},
  \bibinfo{journal}{JCAP} \textbf{\bibinfo{volume}{0711}}, \bibinfo{pages}{011}
  (\bibinfo{year}{2007}), \eprint{0705.0116}.

\end{thebibliography}

\end{document}